\documentclass[letter]{aa}  

\usepackage{graphicx}

\usepackage{txfonts}

\begin{document}

   \title{Unlocking the secrets of the midplane gas and dust distribution \\ in the young hybrid disc HD~141569}

   \author{J. M. Miley \inst{1}, O. Pani\'c \inst{1}, M. Wyatt \inst{2}, G. M. Kennedy \inst{3,4}}

   \institute{School of Physics \& Astronomy, University of Leeds,
              Woodhouse Lane, Leeds, LS2 9JT, UK\\
              \email{py12jm@leeds.ac.uk}
         \and 
         	 Institute of Astronomy, University of Cambridge, Madingley Road, Cambridge, CB3 0HA, UK
         \and
             Department of Physics, University of Warwick, Gibbet Hill Road, Coventry, CV4 7AL, UK
         \and 
         	 Centre for Exoplanets and Habitability, University of Warwick, Gibbet Hill Road, Coventry, CV4 7AL, UK\\
             }

   \date{Submitted 7th May 2018, Accepted 20th June 2018}

 \abstract
 {\object{HD~141569} is a pre-main sequence star with a disc uniquely placed between protoplanetary and debris discs, similar to the older `hybrid' type discs.}
 {This work aims to place the mass and spatial structure of the disc midplane in the context of the debris, hybrid and protoplanetary discs.}
 {We observed HD~141569 with ALMA in 1.3~mm continuum and $^{13}$CO (2-1). This is the first detection and image of the optically thin gas emission from the midplane of this disc.}
 {In continuum emission, we detect a combination of an unresolved central peak and a ring of millimetre emission at 220$\pm$10~au, slightly interior to one of the rings discovered in scattered light. The minimum dust mass of the ring is 0.13$\pm$0.02~M$_{\oplus}$ while the unresolved millimetre peak at the stellar location is predominantly thermal emission due to a minimum of 1.2$\pm$0.2~M$_{\oplus}$ of dust. $^{13}$CO is distributed asymmetrically around the stellar position with a peak at 1$\farcs$1 distance and a P.A. of -33$^\circ$. The gas is detected as far as 220$\pm$10~au, a radial separation the same as that of the mm ring. Assuming optically thin emission and standard ISM abundances, we use our $^{13}$CO data to derive the gas mass in the disc of (6.0$\pm$0.9) $\times 10^{-4}~$M$_\odot$. Comparison to published $^{12}$CO data shows that $^{12}$CO is optically thick, explaining why estimates based on $^{12}$CO underestimated the gas mass.}
 {} 

   \titlerunning{Midplane gas and dust in young hybrid disc HD~141569}
   \keywords{}

   \maketitle


\section{Introduction}

Understanding the evolution from protoplanetary to debris discs \citep{Dominik2002FromDisks,Wyatt2015FiveDisk} requires an understanding of their structures. How do the many gaps, dust traps and asymmetries seen in protoplanetary discs evolve into the one or two dusty debris rings, and what is the role of gas in this process? The `hybrid' discs \citep{Pericaud2017}, are believed to probe this step in evolution because of detectable amounts of gas in the disc despite otherwise appearing to have evolved beyond the protoplanetary disc limit in dust mass. 

HD~141569 is the only known hybrid disc amongst pre-main sequence stars to date, a B9-A0 type, 5~Myr old pre-main sequence Herbig star at a GAIA distance of 111$\pm$1~pc \citep{Weinberger2000StellarDisk,Lindegren2016Gaia1}. \footnote{We have scaled previously reported length scales to the new GAIA distance for the source.}

The SED and IR excess luminosity of 0.0084 \citep{Sylvester1996OpticalSystems} were the first indicators of this unique status. HD~141569 has an SED which appears like a scaled-down version of a protoplanetary disc SED, clearly dominated by multiple radial (temperature) components, discerning it from the known debris and hybrid discs, as both these types show at most two radial (temperature) components \citep[e.g., 49 Ceti, HD~131835,][]{Roberge2013HERSCHELDISK,Moor2015DISCOVERYDISKS,Kennedy2014DoBelts}. The dust mass of 0.7~M$_{\oplus}$ \citep{Sandell2011ASTARS,Panic2013FirstDiscs} places HD~141569 between the protoplanetary and debris discs regimes \citep{Wyatt2015FiveDisk}. First detected by \citet{Dent2005COStars}, the CO gas extends up to 200~au \citep{Flaherty2016RESOLVEDDISK,White2016ALMADISK} and is nested inside the two outermost rings detected by Hubble Space Telescope NICMOS at 234 and 388~au  \citep{Augereau1999HST/NICMOS2Disk,Weinberger1999THENICMOS,Biller2015TheDisc} . Reported gas masses range between 4.5$\times 10^{-6}$~M$_\odot$ derived from $^{12}$CO \citep{White2016ALMADISK} and 5$\times 10^{-4}$~M$_\odot$ derived from SED model fits to Herschel line observations \citep{Thi2014AstrophysicsModeling}. None of these estimates employed optically thin CO isotopologue emission, which until now remained undetected in this disc. The estimated gas mass range for HD 141569 is below the typical protoplanetary disc gas masses of $10^{-3}$~M$_\odot$ to $10^{-1}$~M$_\odot$ but above the hybrid discs gas masses which are mainly <$10^{-6}$~M$_\odot$ \citep{Pericaud2017}. 

Insight into the innermost regions comes from SED fitting, with an inner radius set at 30~au \citep{Marsh2002MID-INFRARED141569}, and CO ro-vibrational emission placing gas as close as 11~au from the star \citep{Goto2006InnerLines}. These regions have recently been imaged with SPHERE on the Very Large Telescope (VLT), detecting a series of concentric, but discontinuous, scattered light `ringlets' at separations of 45, 61 and 89~au \citep{Perrot2016DiscoveryVLT/SPHERE}. Inside these rings is a dust disc component within 0\farcs2, detected in 8.2$\mu$m imaging with VLT VISIR instrument \citep{Perrot2016DiscoveryVLT/SPHERE}. A central, unresolved component of millimetre continuum emission peaks at the stellar location \citep{Marsh2002MID-INFRARED141569,White2018ALMASystem}.   
Beyond the gas-rich region are the two scattered light rings at 234 and 388~au, with inclination and PA similar to that of the inner rings seen with SPHERE \citep{Perrot2016DiscoveryVLT/SPHERE}, but showing more complex structures and spirals, possibly due to planets within the disc \citep{Wyatt2005SpiralDisk} or a fly-by \citep{Reche2009InvestigatingSystem}. 

In this letter we show to what extent the midplane gas and dust of HD~141569 exhibit protoplanetary and debris disc characteristics, in the context of increasing evidence for a `hybrid' class of discs. 

\section{Observations and Results}

Our ALMA observations in Band 6 were taken on 2016, May 16 (PI Pani\'c, ID 2015.1.01600.S). 
39 antennas of 12~m diameter were employed in a configuration providing baseline lengths ranging from 16~m to 640~m. The velocity resolution is 0.166~km/s for $^{13}$CO and 0.333~km/s for C$^{18}$O, providing optimal sensitivity to the tenuous and previously undetected midplane gas emission. The 
total effective continuum bandwidth was 4~GHz. 
The data were flagged and calibrated following the ALMA-pipeline data reduction. 
Titan was used as the flux calibrator, J1550+0527 and J1733-1304 were used as the bandpass and phase calibrators respectively. \footnote{In the analysis of our data we use the available $^{12}$CO(3-2) data published in \citet{White2016ALMADISK}.}
Using CASA 4.7.1, we subtracted the continuum emission in the uv-space. 
Images were created using the clean algorithm \citep{Rau2011AInterferometry}; as we have no a priori knowledge of the distribution of $^{13}$CO or indeed of the dust emission, we adopt natural weighting to prioritise the collection of signal, rather than the increased resolution that could be achieved with other baseline weightings. This produced a synthesized beam size of 0$\farcs$65 (72~au) for the continuum and 0$\farcs$71 (79~au) for $^{13}$CO and C$^{18}$O. 

Rms noise level in the continuum image is 0.11~mJy/beam, and in the $^{13}$CO and C$^{18}$O channel maps rms was 0.04 and 0.03~Jy km/s/beam respectively. In the continuum imaging an unresolved point source of 1.7~mJy is detected. $^{13}$CO does not show a firm detection in the individual channel maps, but the integrated intensity map over the known range of velocities in which $^{12}$CO is seen \citep{Flaherty2016RESOLVEDDISK,White2016ALMADISK} shows a 6$\sigma$ detection in $^{13}$CO.
C$^{18}$O is not detected >3$\sigma$ in either the channels or the integrated intensity map.

\begin{figure}
\centering\includegraphics[width=0.9\linewidth]{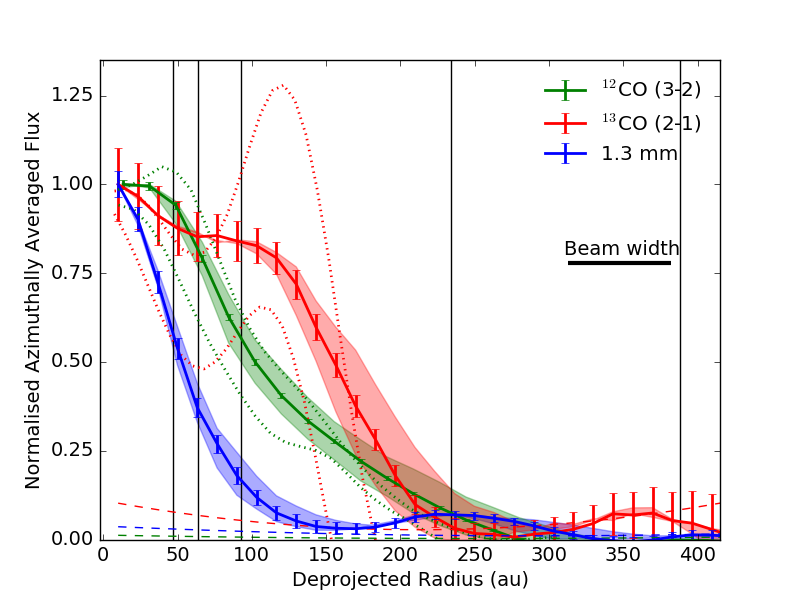}
\caption{Azimuthally averaged flux profile following the deprojection of coordinates from the 1.3~mm image of HD 141569 (blue), $^{12}$CO (green) and $^{13}$CO (red). Vertical black lines show the radial locations of rings identified in scattered light. Peak intensities for 1.3~mm continuum, $^{12}$CO and $^{13}$CO are 1.74 mJy/beam, 1.06 and 0.25 Jy km/s/beam respectively. Coloured dotted lines show NW (upper) and SE (lower) cuts of the $^{12}$CO and $^{13}$CO maps along the position angle of the $^{13}$CO peak.The shaded regions show the variation of the profile when changing inclination of the source by +/-10$^{\circ}$. Error bars are calculated as described in the text, the coloured dashed lines represent the noise level after averaging for each curve.}
\label{fig:azi_avg_cont}
\end{figure}

We perform azimuthal averaging in radial bins of 0$\farcs$12 for the $^{13}$CO and 1.3mm maps, and 0$\farcs$15 for $^{12}$CO. Neither the strongly asymmetric $^{13}$CO nor the unresolved central 1.3mm emission can provide reliable constraints on the position angle and inclination of the disc, so for deprojection of coordinates we adopt the values of the $^{12}$CO disc, PA=-9$^{\circ}$ and inclination=55$^{\circ}$ \citep{White2016ALMADISK}. Fig. \ref{fig:azi_avg_cont} shows the averaged profiles. At 220~au we detect a thin concentration of millimetre grains concentrated into a ring and detect $^{13}$CO(2-1) emission as far as 220$\pm$10~au. 

To derive uncertainties for the radial flux profiles we azimuthally average emission free-regions of the image that are sufficiently far from the central emission but close enough to have high beam efficiency. The mean value across emission-free regions from different parts of the image gives a flux profile that describes the background noise in the image as a result of the averaging process. This method is more robust than simply taking the variation in an elliptical bin and does not rely on any predefined disc models. The resultant noise curves are shown in Fig 1. for each tracer, where the outer belt is detected at $\sim$6$\sigma_{\rm mm}$ in contiguous radial bins (comparing the 1.3~mm curve to the blue dashed noise line), with a peak at 220$\pm$10~au. 

\section{Discussion}

\subsection{Dust emission}

The comparison of the radial profile of 1.3~mm flux with the known disc structures shows that some of the flux in the central unresolved component may be arising from the rings detected by SPHERE \citep{Perrot2016DiscoveryVLT/SPHERE}. We calculate a minimum dust mass for the inner disc by assuming optically thin emission, using $M_{dust}= \frac{F_\nu~d^2}{\kappa_\nu~B_\nu(T)}$, taking a midplane disc temperature of 20~K, an opacity, $\kappa_\nu$, corresponding to mm size grains of 1.15~cm$^{2}$g$^{-1}$ \citep{Draine2006OnDisks} and a distance to the star, d, of 111~pc. The minimum dust mass for the inner disc is calculated to be (3.6$\pm 0.5) \times 10^{-6}$ M$_{\odot}$ = 1.2$\pm$0.2~M$_\oplus$. We have assumed in this calculation that all emission is thermal. A maximum opacity has been chosen to derive a minimum dust mass, the uncertainties correspond to a $\pm$15\% flux calibration uncertainty applied to all ALMA observations. This mass depends on disc temperature, 20 K is a good assumption for cool material in the midplane and outer disc, increasing the assumed midplane temperature to 50 K would give a mass of 0.04~M$_\oplus$.   

An alternative explanation for the origin of this flux may be free-free emission, rather than thermal emission originating from rings of dust within the inner disc. In fact, a typical mm contribution from free-free emission in Herbig discs at a similar distance to HD~141569, is a few mJy \citep[e.g.,][]{Wright2015ResolvingATCA}, similar to the 1.7~mJy integrated flux we measure for the unresolved peak. 
Free-free emission has a characteristic spectral slope of -0.1 for optically thin emission, considerably lower than that of thermal continuum emission in this regime and so its effect may be seen in the spectral index. 
Protoplanetary discs typically show $\alpha_{\rm mm} \approx 2-3$, transition discs tend to show slightly higher values than this \citep{Pinilla2014MillimetreRadius}, whilst debris discs have $\alpha_{\rm mm} \approx 2.5-3$ \citep{MacGregor2016CONSTRAINTSDISKS}. ALMA and VLA observations derive a spectral index for the disc of $\alpha_{\rm mm}$ = 1.63 \citep{MacGregor2016CONSTRAINTSDISKS,White2018ALMASystem}, with which our 1.3~mm integrated flux of 1.8~mJy is consistent. $\alpha_{\rm mm}$ calculated between ALMA observations at 870$\mu$m and 2.9~mm also give a steeper, but still relatively low value of 1.81 \citep{White2018ALMASystem}. VLA observations show a significant variability in the continuum flux of the disc, varying by a factor of 2-3 on timescales of 10s of minutes \citep{White2018ALMASystem}. Fluxes measured on integration times shorter than the variability timescale may therefore only offer a snapshot of flux emission. To make a conservative estimate of possible free-free contribution we extrapolate from  VLA flux of 54~$\mu$Jy \citep{MacGregor2016CONSTRAINTSDISKS} using a characteristic spectral slope for optically thin free-free emission; from this estimate the free-free flux at 1.3mm would be just 3\% of the total integrated flux we measure, meaning the flux we detect is predominantly thermal, potentially with small free-free contribution of order 10s of $\mu$Jy at most. 

\begin{figure}
\centering\includegraphics[width=0.82\linewidth]{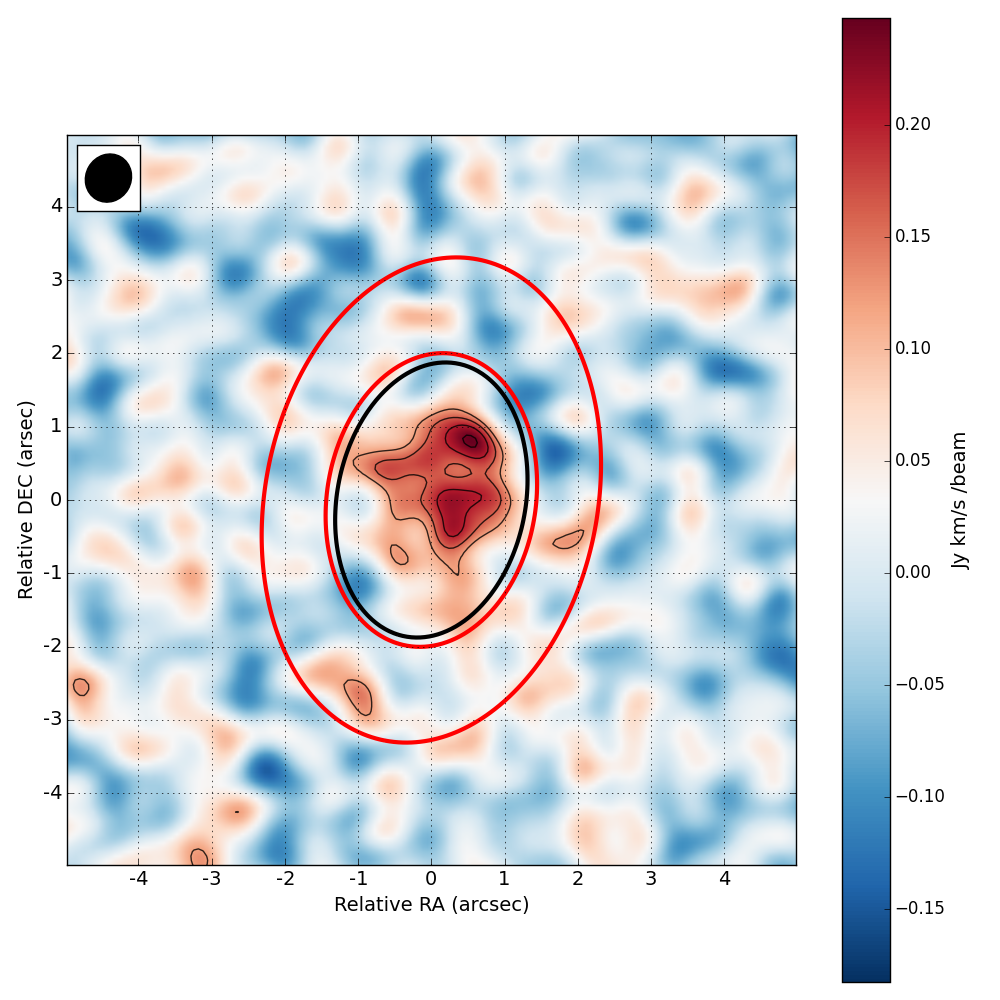}
\caption{$^{13}$CO emission is shown in the colour map and contoured at levels of (3,4,5,6)$\times\sigma_{13}$, with $\sigma_{13}$=0.040 Jy km/s /beam. The red ellipses show the position of the HST rings with parameters from \citet{Biller2015TheDisc}. The black ellipse shows the location of the mm ring detected in the 1.3~mm continuum image.}
\label{fig:13co}
\end{figure}

In Fig. \ref{fig:azi_avg_cont} we see an increase in measured flux which peaks at a deprojected radial separation of 220$\pm$10~au. The ring is detected interior to scattered light ring at 234~au \citep{Biller2015TheDisc} as is demonstrated by the ellipses in Fig. \ref{fig:13co}. Taking the flux from the peak of the ring in Fig. \ref{fig:azi_avg_cont} and assuming the same temperature and opacity as before, we can calculate a minimum dust mass of 0.13$\pm$0.02~M$_\oplus$, firmly in the debris disc regime < 1~M$_\oplus$. We can place an upper limit on the outermost scattered light ring at 388 au from our non-detection in the continuum. Making the same assumptions as before, the maximum mass of the outer ring at which it could still avoid detection in azimuthal averaging is $\sim$0.11$\pm$0.02~M$_\oplus$.


The position of the millimetre ring relative to the gas disc hints at potential formation mechanisms. A continuum ring at the edge of the gas disc is a morphology that can be compared with models of a primordial, but depleted gas disc. In this case radiation pressure overcomes radial drift to push dust particles of <100~$\mu$m outward to the edge of the gas disc \citep{Takeuchi2001DUSTDISKS}. 
The ring emission we detect is likely to come from larger grain sizes; millimetre continuum is emitted most efficiently by a narrow range between 1mm and 1cm \citep{Takeuchi2005AttenuationAccretion}. If blow-out from radiation pressure formed the ring, the small grains collecting up at the edge of the gas disc would have to collide and grow into mm-sized grains. This seems unlikely; some small grains will acquire an eccentricity from this process, resulting in collisions that would occur at higher velocity leading to fragmentation rather than grain growth \citep[see Fig. 7]{Takeuchi2001DUSTDISKS}. Given our results, a more detailed application of this model to HD~141569 is required, but the level of necessary grain growth seems a tall order. 
An alternative scenario might be the sculpting of grains into a ring as a result of an unseen planet or perhaps a flyby. Two M dwarfs in a binary system are located less than 9\farcs0 away from HD~141569 and previous studies of the disc considering a flyby have been able to reproduce global disc features such as the wide gap in the dust and spiral structures  \citep{Augereau2004StructuringDisk,Ardila2005A141569A,Reche2009InvestigatingSystem}. The caveat on this explanation is that the statistical likelihood of such a flyby is still low and the relationship, both present and past, between HD~141569 and the binary is not completely understood \citep{Reche2009InvestigatingSystem}.

\begin{figure}
\centering\includegraphics[width=0.82\linewidth]{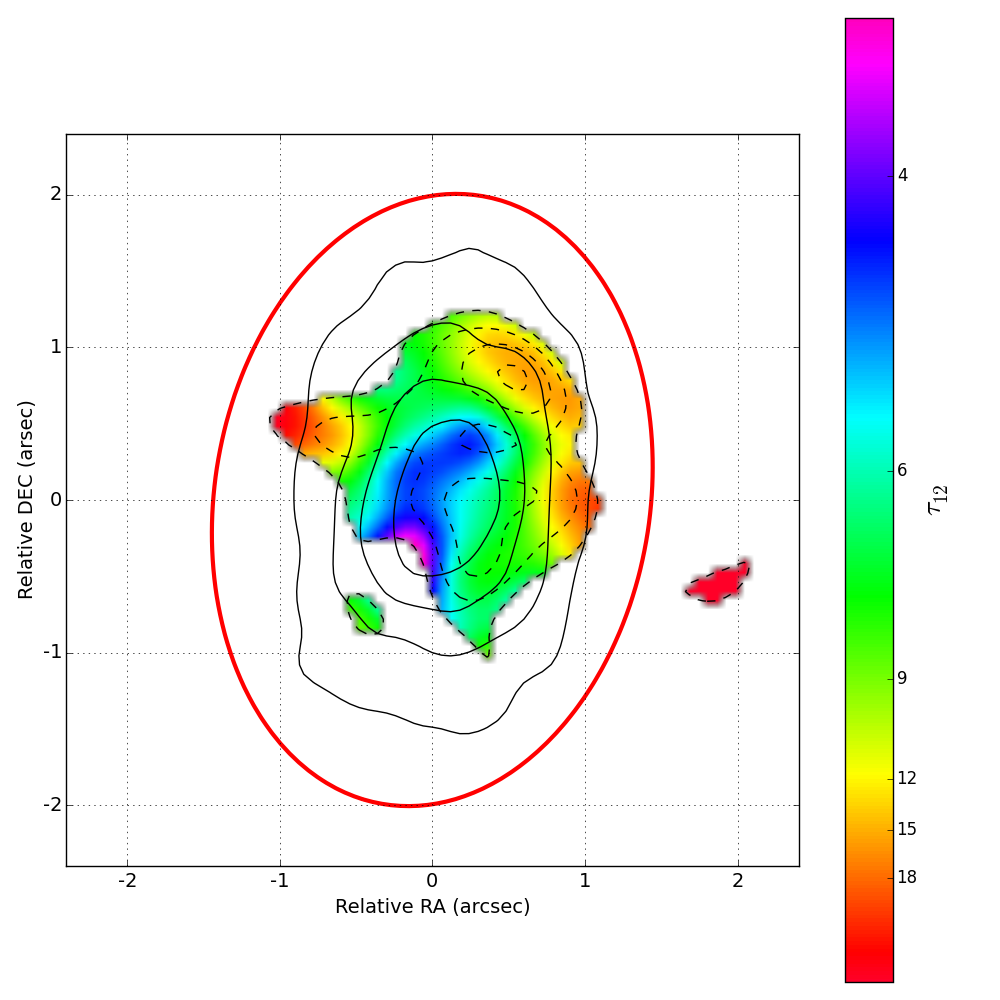}
\caption{Map of the optical depth of $^{12}$CO, the colour map is plotted for regions of the disc where we detect $^{13}\rm CO>3\sigma_{13}$ . Overlaid in solid black contours of (10,20,30,40)$\times\sigma_{12}$ ($\sigma_{12}$ = 0.064 Jy km/s /beam) is the $^{12}$CO integrated intensity map convolved with the larger beam and dashed black contours show $^{13}$CO as in Fig \ref{fig:13co}.}
\label{fig:tau}
\end{figure}

\subsection{Gas emission}

The $^{13}$CO emission is strongly asymmetrical (Fig~\ref{fig:13co}). The emission peak is measured at 6$\sigma_{13}$ off centre from the stellar position, located at a deprojected separation of 1$\farcs$1 (120$\pm$10~au) at a position angle of -33$^\circ$. Whilst at the same separation towards the South East direction, emission is measured <2$\sigma_{13}$. The emission we detect from $^{13}$CO is roughly co-spatial with the scattered light emission of the eccentric dust ring found at a separation of 89~au \citep[][ their Fig. B3]{Perrot2016DiscoveryVLT/SPHERE}. Intriguingly, the ring is brighter in scattered light at the opposite side of the disc.
$^{13}$CO is detected over a considerably larger radial range than the inner edge of the gas disc at 11~au seen in CO rovibrational emission \citep{Goto2006InnerLines}, than the inner edge of the dust disc of 30~au derived from the SED \citep{Marsh2002MID-INFRARED141569}, and than the asymmetry seen in $^{12}$CO channel maps \citep{White2016ALMADISK}. 

$^{13}$CO emission is radially the brightest within the same range of stellocentric distances in which the scattered light rings were found, at 45, 61 and 89~au \citep{Perrot2016DiscoveryVLT/SPHERE}, and co-spatial with the HST detection of scattered light between 0$\farcs$4 and 1$\farcs$0 \citep{Konishi2016DISCOVERYA}. 
Through azimuthal averaging (Fig~\ref{fig:azi_avg_cont}) we find that the largest radial distance $^{13}$CO is detected to is 220$\pm$10~au. This separation is coincident with the location of the ring we detect in millimetre emission, with both interior to the 234~au scattered light ring. 

Such an asymmetry in molecular gas has been seen in the debris disc Beta Pic \citep{Dent2014MolecularDisk.}, which may have resulted from giant collisions \citep[e.g.][]{Jackson2014DebrisRadii}, but which is more likely to be due to the collisional destruction of planetesimals trapped in resonance with a planet \citep{Wyatt2003ResonantSystem,Matra2017ExocometaryObservations}. This scenario would make the gas secondary and so may be different to HD~141569 if the gas in the latter is primordial; the large radial extent of the gas is more typical of a primordial origin, but the nature of the gas in the disc is yet to be confirmed. 
Younger, protoplanetary, discs can also exhibit strong azimuthal asymmetry that is typically strongest in the dust emission, e.g. Oph IRS 48, MWC 758 \citep{vanderMarel2013ADisk,Boehler2017TheCavity}. `Lop-sided' discs form as a result of dust traps that depend upon the surface density and turbulence of the disc \citep{Pinilla2012TrappingDisks,Birnstiel2013LopsidedDisks}, making it an unlikely mechanism in the less massive hybrid and debris discs.

To determine the optical depth of $^{12}$CO, we calculate the spatial distribution of $\tau_{^{12}\rm CO}$ by using the $^{12}$CO and $^{13}$CO images. We convert the higher transition $^{12}$CO (3-2) emission into the minimum expected $^{12}$CO (2-1) emission using a  $^{12}$CO(2-1)/$^{12}$CO(3-2) ratio (1.08) obtained from the RADEX online tool for the disc temperature of 50K \citep{vanderTak2007ASpectra}\footnote{Available at home.strw.leidenuniv.nl/moldata/radex.html}. Using lower temperatures does not change this ratio by more than a factor of 2 and ensures we use the minimum amount of CO(2-1), therefore calculating the minimum optical depth. We convolve both the $^{12}$CO and $^{13}$CO images with the synthesized beam of the $^{13}$CO image in order to ensure the images have the same effective smoothing. Under the assumption that both lines have the same excitation temperature, the ratio of  $^{12}$CO(2-1) and $^{13}$CO(2-1) emission is then used to compute the optical depths of both lines, following the method used in, e.g. \citet{Schwarz2016THEISOTOPOLOGUES}, and assuming the ratio of optical depths to follow the ISM abundance of 77. The obtained map (Fig. \ref{fig:tau}) shows that $\tau_{^{12}\rm CO}$ ranges from 3 to 45 (Fig.~\ref{fig:tau}). $\tau_{^{13}\rm CO}$ is therefore in the range from 0.04 to 0.58.

Fig.~\ref{fig:tau} confirms $^{12}$CO (2-1), and therefore also (3-2), to be optically thick throughout the disc, while the $^{13}$CO is not. In central regions $\tau_{^{12}\rm CO}$ $\sim$ 2 where $^{12}$CO is strongest, reaching as high as 25 in outer regions where $^{13}$CO is detected strongly. Once $^{12}$CO becomes optically thick, it ceases to trace mass in the midplane because emission is coming from upper layers of the disc at greater scale heights. 
This may also explain why the asymmetry of the disc is so prominent in the $^{13}$CO map, but not in the $^{12}$CO; the optically thick $^{12}$CO emission is dominated by the temperature of the emitting layer, which increases closer to the star, so a combination of temperature and emitting surface area are responsible for the brightness of emission, rather than the fractional changes in gas density. However the optically thin $^{13}$CO traces density more faithfully and provides a better indicator of the spatial distribution of gas in the disc, highlighting any asymmetries which may be hidden by temperature and surface area effects in $^{12}$CO.
The radial profiles shown in Figure \ref{fig:azi_avg_cont} illustrate this effect: the azimuthally averaged flux drops sharply after the peak at around $\sim$30~au for $^{12}$CO, but the $^{13}$CO emission is still >75\% of the peak value until after 100~au. This is clearer in the NW/SE slices in Fig. \ref{fig:azi_avg_cont} taken in the direction of the $^{13}$CO peak. In green, the $^{12}$CO peaks much closer to the star where temperature effects are more prominent ($\sim$ 30~au in the NW slice), whilst the $^{13}$CO peaks much further out in the disc at a deprojected distance of $\sim120$~au. The slices also illustrate the magnitude of the asymmetry in the midplane gas, with a very strong $^{13}$CO peak towards the North West that is considerably stronger than the average flux at that separation.
This demonstrates the importance of using CO isotopologues, as done here, to investigate the disc midplane, and the gas density distribution within it.

Having shown that the $^{13}$CO emission is optically thin, we derive a lower limit on the mass of the gas disc using 
\begin{equation}
M_{gas} = \frac{4 \pi}{h \nu_{21}} \frac{F_{21} m d^{2}}{A_{21} x_2} \Bigg[\frac{\rm H_2}{^{12}\rm{CO}}\Bigg] \Bigg[\frac{^{12}\rm{CO}}{^{13}\rm{CO}}\Bigg] 
\end{equation}

where F$_{21}$ is the integrated line flux from the $J$=2--1 transition of $^{13}$CO, m is the mass of the CO molecule, A$_{21}$ the Einstein coefficient for the transition, $\nu$ is the frequency at which the transition occurs and x is the fractional population of the upper level. We take the midplane temperature to be 20K and assume ISM abundances of $^{12}$CO/$^{13}$CO = 77 and $^{12}$CO/H$_2$=10$^{-4}$ \citep{Wilson1994AbundancesMedium} to give a total gas disc minimum mass of (6.0$\pm0.9) \times 10^{-4}$M$_\odot$. Uncertainties are due to the ALMA 15\% flux calibration uncertainty. Like the dust mass, this value depends on disc temperature, for example adopting T=50K would give (9.0$\pm 1.0)\times 10^{-4}$M$_\odot$. Sequestration of CO into icy bodies or conversion into more complex molecules can reduce gas-phase CO in the disc, decreasing the gas mass calculated by this method. Our minimum mass is two orders of magnitude larger than that calculated using the same method but with $^{12}$CO as the mass tracer, $\sim4.5 \times 10^{-6}$ M$_\odot$ \citep{White2016ALMADISK}. 
SMA observations CO(1-0) emission gives a gas disc mass 1.05$\times 10^{-4}$M$_\odot$, comparable to the mass we derive here, however MCMC modelling of the data gave a slightly lower value of 3.8$\times 10^{-5}$M$_\odot$ \citep{Flaherty2016RESOLVEDDISK}. \citet{Thi2014AstrophysicsModeling} also find a similar order of magnitude by fitting Herschel gas line observations in the far-IR to simultaneous radiative transfer and chemical modelling, giving a range of 2.5-5 $\times 10^{-4}$M$_\odot$. Each of these methods assume $^{12}$CO/H = $10^{-4}$.

$^{12}$CO therefore appears to be underestimating disc mass, agreeing with our confirmation of the tracer's optical thickness. A similar scenario was found in 30 Myr old hybrid disc HD~21997; \citet{Kospal2013ALMA21997} calculate a CO gas mass in the disc from the optically thin C$^{18}$O line that was 2 orders of magnitude larger than that previously calculated assuming optically thin $^{12}$CO emission \citep{Moor2011MOLECULARDISKS}.

Our observations do not detect C$^{18}$O, but from the rms noise in the integrated image we can derive an upper limit on the C$^{18}$O gas mass and a corresponding estimate of the disc gas mass. Rms in the C$^{18}$O image with natural weighting is 25~mJy/beam, yielding a total gas mass estimate of 1.4$\times$10$^{-4}$~M$_\odot$ assuming T=20~K, ISM abundances and isotopic ratios. This value is higher than that traced by the optically thick $^{12}$CO, but much lower than the estimate based on the $^{13}$CO. In fact, the flux ratio between $^{13}$CO and the upper limit on C$^{18}$O flux is 31, significantly higher than the typical ISM abundance of 8 \citep{Wilson1994AbundancesMedium}. This implies that C$^{18}$O is under abundant by at least a factor 3.8. An explanation for this may be the isotope-selective photo-dissociation, known to deplete the less abundant isotopologues such as C$^{18}$O in discs \citep{Visser2009TheDisks,Miotello2016DeterminingObservations}. The higher column density of the more abundant isotopologue, in this case $^{13}$CO, means that it can self-shield from photodissociation more effectively.
It is important to note that isotope-selective photodissociation cannot be the cause of the differences between the gas mass estimates based on $^{12}$CO and $^{13}$CO, as it would have the opposite effect: The mass based on 
$^{12}$CO should be higher in such case. In case of optically thick $^{12}$CO emission, as here, only the comparisons between the optically thin isotopologue lines allows us to assess the selective photodissociation effects, again stressing the importance of observing these species.

\subsection{Nature of the disc}

In this section we compare our new results on disc structure and mass of gas and dust to some similar systems. 

Known hybrid discs generally have disc gas masses\footnote{We assume here $^{12}$CO/H$_2$=10$^{-4}$ to convert CO masses to total disc masses in order for ease of comparison. For individual discs this ratio may be less appropriate than others.} derived from CO observations of order one Earth mass \citep[e.g. 49 Ceti 6.6$\times$10$^{-6}$~M$_\odot$ , HD 131835 1.3$\times$10$^{-5}$~M$_\odot$,][]{Hughes2008ACeti,Moor2015DISCOVERYDISKS}. HD~141569 has a particularly massive gas disc in comparison \citep[see e.g. Fig. 6 of ][]{Pericaud2017}. All previous estimates were based on optically thick tracers, which explains why in this work we derive (6.0$\pm0.9$)$\times$10$^{-4}$~M$_\odot$, i.e., two orders of magnitude greater mass. A more direct comparison can be made with hybrid disc HD~21997, where the disc mass derived from ALMA observations of optically thin C$^{18}$O giving 0.8-1.8$\times$10$^{-4}$~M$_\odot$ \citep{Kospal2013ALMA21997} a factor of a few less than HD~141569. Hybrid discs in general may be hosting more gas than traced by CO and its isotopologues, not only due to optical depth but also if CO has been depleted with respect to H$_2$. 

The minimum dust mass 1.2$\pm$0.2~M$_\oplus$ corresponding to the unresolved component is greater than the dust masses of HD~131835, \citep[0.47~M$_\oplus$][]{Moor2015DISCOVERYDISKS}, HD~21997 \citep[0.09~M$_\oplus$][]{Moor2013ALMA21997} and 49 Ceti \citep[0.2~M$_\oplus$][]{Holland2017SONS:Submillimetre}.  

The millimetre ring at 220~au has a minimum dust mass of 0.13$\pm$0.02 ~M$_{\oplus}$. This mass and the scattered light ring seen at a slightly larger distance are both comparable to typical debris discs. Debris systems show evidence of rings containing millimetre sized grains at large radii outside an inner debris component like HD~181327 \citep{Marino2016ExocometaryRing} and HD~107146 \citep[Marino et al., submitted.]{Ricci2015ALMA107146}, perhaps the nesting of structure within a larger ring is a more common component than we currently anticipate.

\section{Conclusions}
We have detected an asymmetric distribution of midplane gas in the young hybrid disc HD 141569, traced by optically thin $^{13}$CO emission. We also detect a faint mm ring of emission at 220$\pm$10~au, providing a midplane counterpart to the ring detected in scattered light, where these small dust grains originate. We derive the minimum dust mass in the ring of 0.13$\pm$0.02 M$_\oplus$, around 10\% of the total flux from the system. The bulk of the millimetre emission, known to be variable, coincides with the location of the star and is unresolved, placing an upper limit on its radial extension of 72~au. This emission traces a minimum of 1.2$\pm$0.2 M$_\oplus$ of dust, but may have a free-free contribution of $\le$3\%. From our observations we are able to place an upper limit on the outermost ring of 0.11$\pm$0.02~M$_\oplus$. 

The gas disc is interior to this ring at 220~au. From the integrated $^{13}$CO flux we derive a gas mass for the disc of (6.0$\pm$0.9)$\times 10^{-4}$M$_\odot$; a mass $\sim$ 2 orders of magnitude greater than that derived using $^{12}$CO. We have shown that the reason for this is the high optical depth of $^{12}$CO, rendering any gas mass value derived from its emission a large underestimate. Non-detection of C$^{18}$O shows that this isotopologue is about four times less abundant than we would expect, based on the standard isotopic ratio between $^{16}$O and $^{18}$O, and we propose the isotope-selective photodissociation as the culprit.

\textit{At the time of completion of the work on our data, the detection of the continuum ring was presented in a publication by others \citep{White2018ExtendedALMA}. We nonetheless present the analysis of our entire dataset, including both gas lines and the dust.}

\begin{acknowledgements}
We thank the anonymous referee for a valuable review. The work of J.~M. is supported through the University Research Scholarship of the University of Leeds. The work of O.~P. is funded by the Royal Society through the Dorothy Hodgkin Fellowship. G.~M.~K. is supported by the Royal Society as a University Research Fellow. 
\end{acknowledgements}

\bibliographystyle{aa}
\bibliography{Mendeley.bib} 
\end{document}